\documentclass{ws-ijmpb}

\begin{document}

\markboth{Stefano Longhi}
{Accelerated and Airy-Bloch oscillations}

\catchline{}{}{}{}{}
%
%%%%%%%%%%%%%%%%%%%%%%%%%%%%%%%%%%%%%%%%%%%%%%%%%%%%%%%%%%%%%%%%%%%%

\title{Accelerated and Airy-Bloch oscillations}

\author{STEFANO LONGHI}

\address{Dipartimento di Fisica e Istituto di Fotonica e Nanotecnologie del Consiglio Nazionale delle Ricerche, Politecnico di Milano, Piazza Leonardo da Vinci 32\\
Milan, 20133, Italy\\
stefano.longhi@fisi.polimi.it}

\maketitle

%\begin{history}
%\received{Day Month Year}
%\revised{Day Month Year}
%\accepted{(Day Month Year)}
%\comby{(xxxxxxxxxx)}
%\end{history}

\begin{abstract}
A quantum particle subjected to a constant force undergoes an accelerated motion following a parabolic path, which differs from the classical motion just because of wave packet spreading (quantum diffusion). However, when a periodic potential is added (such as in a crystal) the particle undergoes Bragg scattering and an oscillatory (rather than accelerated) motion is found, corresponding to the famous Bloch oscillations. Here we introduce an exactly-solvable quantum Hamiltonian  model, corresponding to a generalized Wannier-Stark Hamiltonian $\hat{H}$, in which a quantum particle shows an intermediate dynamical behavior, namely an oscillatory motion superimposed to an accelerated one. Such a novel dynamical behavior is referred to as {\it accelerated Bloch oscillations}. Analytical expressions of the spectrum, improper eigenfunctions and propagator of the generalized Wannier-Stark Hamiltonian $\hat{H}$ are derived. Finally, it is shown that acceleration and quantum diffusion in the generalized Wannier-Stark Hamiltonian are prevented for Airy wave packets, which undergo a periodic breathing dynamics that can be referred to as {\it Airy-Bloch oscillations}.
\end{abstract}


\section{Introduction}
Stark Hamiltonians play a central role in several fields of physics, including quantum mechanics, condensed matter physics and optics \cite{r1,r2,r3}. In non-relativistic quantum mechanics, the simplest case is that of a quantum particle subjected to a constant force $F$, which is described by the Hamiltonian
\begin{equation}
\hat{H}=- \epsilon \frac{\partial^2}{\partial x^2}+Fx \equiv \epsilon \hat{p_x}^2 + Fx
\end{equation}
where the first term of the right hand side of Eq.(1), $ \epsilon \hat{p}_x^2$, is the kinetic energy operator, the second term is the potential of the external force, and $\hat{p}_x=-i \partial_x$ is the momentum operator. As is well known (see e.g. chap.8.1 of Ref.\cite{r2} or Ref.\cite{r4}) the spectrum of $\hat{H}$ is purely absolutely continuous with a set of (improper) eigenfunctions given by shifted Ary functions. Like in the classical limit, a quantum wave packet undergoes a uniformly accelerated motion but spreads owing to quantum diffusion. Indeed, after a change of reference frame (from th rest frame to an accelerated one) and a gauge transformation of the wave function the Stark Hamiltonian (1) is equivalent to the Hamiltonian of a freely moving particle, i.e. with $F=0$ (see, for example, \cite{r5,r6,r7}). A much richer and physically relevant dynamics arises when a periodic potential is added to the external force, i.e. for the Hamiltonian
\begin{equation}
\hat{H}= \epsilon \hat{p}_x^2+V(x)+Fx
\end{equation}
where $V(x)$ is the periodic potential with period $d$. Such a problem was originally studied by Bloch, Zener and Wannier in a series of seminal papers in connection with the motion of an electron in a crystal subjected to a uniform electric field \cite{r8,r9,r10,r11,r11bis}. Under the assumption that a single lattice band is excited and the force is weak to neglect tunneling into other bands of the crystal (Zener tunneling), the quantum motion of the electron can be described by an effective single-band Wannier-Stark Hamiltonian \cite{r3,r12}. Remarkably, owing to Bragg scattering in the lattice a quantum wave packet in the crystal undergoes an oscillatory (rather than uniformly accelerated) motion with period $T_B=2 \pi /(Fd)$, the famous Bloch oscillations (BOs).      
While in natural crystals electronic BOs have never been observed owing to dephasing and many-body effects, BOs have been impressively demonstrated in a number of experiments after the advent of semiconductor superlattices as terahertz
radiation emitted from coherently oscillating electrons \cite{r13,r14}. 
Quantum and classical analogues of BOs  have been
also proposed and observed in a wide variety of different
physical systems, such as in ultracold atoms  and Bose-
Einstein condensates \cite{r15,r16,r17,r18,r19,r20,AP2}, optical waveguide arrays \cite{r21,r22,r23,r24,r25,r26},  photonic
\cite{r27,r28} and acoustic \cite{r29} superlattices, metamaterials \cite{r290}, plasmonic systems \cite{r29bis,r29tris}, and quantum liquids \cite{AP1} to mention  a few. There is a very extended literature on BOs and their applications. Several authors have investigated BOs in presence of lattice disorder and inhomogeneities \cite{r30,r31,r32},  particle interaction and correlation \cite{r33,r34,r35,r36,r37,r38,r39}, inhomogeneous fields \cite{r40}, complex potentials \cite{r41,r42}, time-dependent forces \cite{r43,r44}, BOs of non-classical states of light \cite{r45,r46}, etc. 
As far as applications is concerned, BOs  have  been  used  as  a  tool to  explore  topological properties of band  structures \cite{r47,r48},  to make  precision  measurements  of  gravity \cite{r49},  and  to realize beam splitters for matter waves \cite{r50,r51}. \par
In this work we introduce an exactly-solvable generalized Wannier-Stark Hamiltonian in which the dynamics of a quantum wave packet shows a behavior intermediate between the uniform acceleration of Hamiltonian (1) and the oscillatory BO dynamics of Hamiltonian (2). Namely, an oscillatory dynamics is superimposed to a parabolic motion, an effect that will be referred to as {\it accelerated Bloch oscillations}. The generalized Wannier-Stark Hamiltonian considered in this work has the form\begin{equation}
\hat{H}= \epsilon \hat{p}_x^2 + T( \hat{p}_x)+Fx,
\end{equation}
where $T(q)$ is a periodic function of $q$ with period $ 2 \pi /d$ and zero mean. It differs from Eq.(1) because of the additional term $T (\hat{p}_x)$ to the free-particle kinetic energy operator.  A possible physical implementation of such an Hamiltonian has been recently introduced in optics Ref.\cite{r52}, where  transverse light dynamics in a self-imaging optical cavity is described by an Hamiltonian of the form described by Eq.(1). In Ref.\cite{r52}, the limiting case $\epsilon=0$ was mainly investigated, corresponding to pseudo-periodic BOs dynamics and absence of quantum diffusion. Here we consider the more general case $\epsilon \neq 0$ and derive analytical expressions of the spectrum, eigenfunctions and propagator of the Hamiltonian (3). In particular, we prove rigorously that for $\epsilon \neq 0$ the new phenomenon of {\it accelerated Bloch oscillations} arises: an oscillatory motion with period $T_B= 2 \pi /(Fd)$, i.e. analogous to BOs, is superimposed to the freely-accelerating motion of the wave packet (as if $T=0$ in (3)). Such a result holds for any normalizable wave packet, but can be violated for an initial non-normalizable probability distribution. Interestingly, for an initial condition corresponding to a (non-normalizable) Airy wave packet, eigenfunction of (1), a periodic oscillatory dynamics without acceleration is found, a phenomenon that can be referred to as {\it Airy-Bloch oscillations}.

\section{Generalized Wannier-Stark Hamiltonian}
Let us consider a quantum system described by the one-dimensional Schr\"{o}dinger equation 
\begin{equation}
 i \frac{ \partial  \psi}{\partial t} = \hat{H} \psi(x,t)
 \end{equation}
  with Hamiltonian given by Eq.(3), where 
\begin{equation}
T(q)=\sum_{n=-\infty}^{\infty} T_n \exp(i n d q)
\end{equation}
is a periodic function of period $2 \pi /d$ with zero mean ($T_0=0$) and $p_x= -i \partial_x$ is the momentum operator. For $\epsilon=0$, the Hamiltonian (3) is a  Wannier-Stark Hamiltonian with a periodic kinetic energy operator that is generally introduced in solid-state physics as an effective Hamiltonian to describe single-band electron dynamics in slowly-varying external fields \cite{r3,r12}. In this case  $T(q)$ is the band dispersion curve, and the discretization $x=nd$ ($n=0, \pm1, \pm2, \pm3,...$) at localized Wannier sites must be accomplished \cite{r3,r12}. As is well-known, for $\epsilon=0$ and provided that the discretization $x=nd$ is accomplished, $\hat{H}$ has a pure point spectrum given by an equally-spaced Wannier-Stark ladder with energy separation $\omega_B=Fd= 2 \pi /T_B$; in real space an initially localized wave packet undergoes periodic BOs with period $T_B$. The generalized Wannier-Stark Hamiltonian (3) that we consider in this work differs from the effective Hamiltonian introduced in the theory of electron dynamics in crystals \cite{r3,r12} because (i) the variable $x$ is continuous rather than discrete, and (ii) the term $\epsilon \hat{p}_x^2$  breaks the periodicity of the kinetic energy operator. In such a case, $\hat{H}$ has a pure absolutely continuous spectrum $-\infty < E < \infty$ and the improper (non-normalizable) eigenfunctions $\phi_E(x)$, satisfying the eigenvalue equation 
\begin{equation}
\hat{H} \phi_E(x)=E \phi_E(x),
\end{equation}
 can be analytically determined in terms of series of shifted Airy functions. The explicit expression of $\phi_E(x)$, derived in Appendix A, reads
\begin{equation}
\phi_E(x) =\sum_{n=-\infty}^{\infty}  \rho_n {\rm Ai} \left( \left( \frac{F}{\epsilon} \right)^{1/3} (x-nd-E/F) \right)
\end{equation}
where we have set 
\begin{equation}
\rho_n=\frac{d}{2 \pi \epsilon^{1/3}F^{1/6}} \int_{-\pi/d}^{\pi/d} dq \exp \left\{  i q d n+ \frac{i}{F} \int_0^q d \xi  T( \xi) \right\}.
 \end{equation}
and the following normalization condition holds
\begin{equation}
\langle \phi_{E'}(x) | \phi_E(x) \rangle \equiv \int_{-\infty}^{\infty} dx \phi_{E'}^*(x) \phi_E(x)=\delta(E-E').
\end{equation}
It is interesting to specialize the general result given by  Eqs.(7) and (8) to the following two limiting cases.\\
(i) In the limit $T(q)=0$, i.e. when the Hamiltonian (3) reduces to the standard Stark Hamiltonian (1), one has $\rho_{n}=(\epsilon^{-1/3} F^{-1/6}) \delta_{n,0}$ and thus the improper eigenfunctions (7)  reduce to the usual form (see \cite{r4})
\begin{equation}
\phi_E(x)=\frac{1}{\epsilon^{1/3}F^{1/6}} {\rm Ai} \left( \left( \frac{F}{\epsilon} \right)^{1/3} (x-E/F) \right).
\end{equation}
(ii) In the limit $\epsilon \rightarrow 0$, i.e. when the Hamiltonian (3) has the form of the effective Wannier-Stark Hamiltonian of solid-state physics with a periodic kinetic energy operator \cite{r3,r12}, taking into account that
\begin{equation}
\lim_{ \gamma \rightarrow 0} \frac{{\rm Ai} (x/ \gamma)}{ \gamma} = \delta(x)
\end{equation} 
one obtains
\begin{equation}
\phi_E(x) =\sum_{n=-\infty}^{\infty}  \sigma_n \delta ( x-nd-E/F) 
\end{equation}
where we have set
\begin{equation}
\sigma_n=\frac{d}{2 \pi F^{1/2}} \int_{-\pi/d}^{\pi/d} dq \exp \left\{  i q d n+ \frac{i}{F} \int_0^q d \xi  T( \xi) \right\}.
 \end{equation}
Equations (12) and (13) justify the result previously given in Ref.\cite{r52} and indicate that, even at $\epsilon =0$, the spectrum of $\hat{H}$ is absolutely continuous. Such a result stems from the fact that the variable $x$ here is considered a continuous variable, rather than discretized as in single-band electron transport theory \cite{r12}.

\section{Wave packet dynamics and accelerated Bloch oscillations}

\subsection{Propagator}
The solution to the Schr\"{o}dinger equation (4) with a given initial condition $\psi(x,0)$ can be formally written as
\begin{equation}
\psi(x,t)= \exp(-it \hat{H}) \psi(x,0) = {\hat U}(t) \psi(x,0)= \int_{-\infty}^{\infty} dy \; \mathcal{U}(x,y,t) \psi(y,0). \;\;\;\;\;\;
\end{equation}
The propagator $\hat{U}$, or its kernel $\mathcal{U}(x,y,t)$ entering in Eq.(14), can be determined from the spectral representation of $\hat{H}$, i.e. one has
\begin{equation}
\mathcal{U}(x,y,t)= \int_{-\infty}^{\infty} dE \phi_E^*(y) \phi_E(x) \exp(-iEt).
\end{equation}
Substitution of Eq.(7) into Eq.(15) and after some cumbersome calculations one obtains (see Appendix B for technical details)
{\small
\begin{equation}
\mathcal{U}(x,y,t)=F^{1/3} \epsilon^{2/3} \sqrt{\frac{1}{4 \pi i \epsilon t}}  \exp \left( -iFxt -i \frac{\epsilon F^2 t^3}{3} \right) \sum_{l= -\infty}^{\infty}  G_l(t) \exp\left[ -\frac{\left( y-x+dl- \epsilon F t^2 \right)^2}{4 i \epsilon t} \right] \;\;
\end{equation}}
\noindent where we have set
\begin{equation}
G_l(t) \equiv \sum_{n=-\infty}^{\infty} \rho_n \rho_{n-l}^* \exp(iFtdn)
\end{equation}
and $\rho_n$ are given by Eq.(8). Exact analytical expressions of the function $G_l(t)$, and thus of the propagator (16), can be given in the important case of a sinusoidal shape of $T(q)$, i.e. $T(q)=\kappa \cos(qd)$, which in electronic crystal theory describes a nearest-neighbor tight-binding lattice band. This important and exactly-solvable case is presented in Appendix C [see Eq.(C.6)].

It is interesting to specialize the form of the propagator given by  Eq.(16)  to the following two limiting cases.\\
(i) In the limit $T(q)=0$, i.e. when the Hamiltonian (3) reduces to the standard Stark Hamiltonian (1), one has $G_l(t)=( \epsilon^2 F)^{-1/3} \delta_{l,0}$ and thus
\begin{equation}
\mathcal{U}(x,y,t)= \sqrt{\frac{1}{4 \pi i \epsilon t}}  \exp \left( -iFxt -i \frac{\epsilon F^2 t^3}{3} \right)  \exp\left[ -\frac{\left( y-x- \epsilon F t^2 \right)^2}{4 i \epsilon t} \right] 
\end{equation}
which is precisely the propagator of the Stark Hamiltonian (1) (see for example  Eq.(7.4) in Ref.\cite{r1}).\\
(ii) In the limit $\epsilon=0$,  Eq.(16) shows a singular behavior and the propagator $\mathcal{U}$ can be calculated as a limit of Eq.(16) as $\epsilon \rightarrow 0$. 
Since $\mathcal{U}$ is the kernel of an integral transformation [Eq.(14)], by use of the phase stationary method  as $ \epsilon \rightarrow 0$ one can set
\begin{equation}
\sqrt{\frac{1}{4 \pi i \epsilon t}} \exp \left( -\frac{x^2}{4i \epsilon t} \right) \sim \delta (x)
\end{equation} 
in all terms under the sum in Eq.(16). Using this result, after some straightforward calculations one obtains
\begin{equation}
\mathcal{U}(x,y,t)=F \exp(-iFxt) \sum_{l=-\infty}^{\infty} \left( \sum_{n-\infty}^{\infty} \sigma_n \sigma_{n-l}^* \exp(iFtdn) \right) \delta(x-y-dl), \;\;\;
\end{equation}
where $\sigma_n$ are defined by Eq.(13). 

\subsection{Accelerated Bloch oscillations}
Propagation of an arbitrary initial wave packet $\psi(x,0)$ is determined by the integral transformation (14) with the propagator $\mathcal{U}$ defined by Eq.(16). To study the properties of wave packet evolution, let us distinguish three cases:\\
\\
\begin{figure}[htb]
\centerline{\includegraphics[width=11cm]{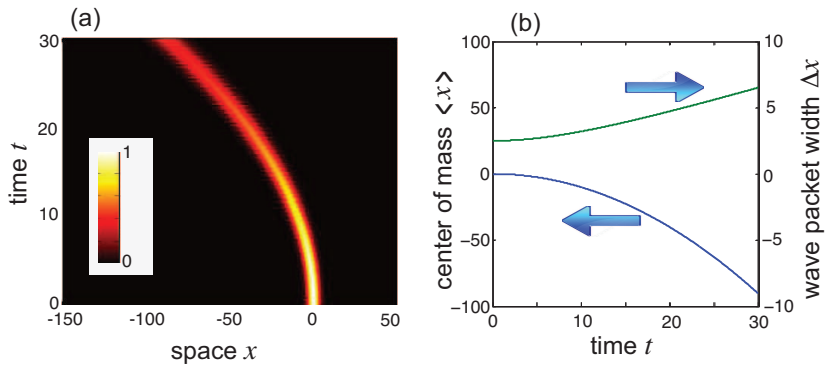}} \caption{Parabolic motion and spreading of a Gaussian wave packet for the Stark Hamiltonian (1). (a) Evolution of the probability density $|\psi(x,t)|^2$ on a pseudocolor map. (b) Evolution of the wave packet center of mass $\langle x(t) \rangle$ and width $\Delta x$. Parameter values are given in the text.}
\end{figure}
(i) {\it First case: parabolic motion.}  This case corresponds to the Stark Hamiltonian (1), i.e. $T(q)=0$. The propagator is given by Eq.(18), which describes particle acceleration and quantum diffusion. An initial wave packet undergoes uniform acceleration and  the center of mass of the wave packet $\langle x (t) \rangle$ follows the parabolic trajectory \cite{note1}
\begin{equation}
\langle x (t) \rangle = \langle x (0) \rangle+ 2 \epsilon \langle \hat{p}_x (0) \rangle t- \epsilon F t^2. 
\end{equation} 
Analytical expression for $\psi(x,t)$ can be given, for example, for an initial Gaussian wave packet distribution. In addition to the parabolic motion, wave packet spreading is observed as a result of quantum diffusion. An example of accelerated Gaussian wave packet is shown in Fig.1. In Fig.1(a) the evolution of the probability density $|\psi(x,t)|^2$ is depicted in a pseudo color map for parameter values $\epsilon=1/2$, $F=0.2$, $T(q)=0$ and for the initial condition $\psi(x,0) \propto \exp(-x^2/w^2)$ with $w=5$. Figure 1(b) shows the evolution of the wave packet center of mass $\langle x(t) \rangle$ and width $\Delta x(t)$ defined by
\begin{eqnarray}
\langle x(t) \rangle & = & \int_{-\infty}^{\infty} dx \; x |\psi(x,t)|^2 \\
\Delta x(t) & = & \left\{ \int_{-\infty}^{\infty} dx (x- \langle x(t) \rangle)^2 |\psi(x,t)|^2 \right\}^{1/2}.
\end{eqnarray}
Note that the wave packet center of mass follows the parabolic trajectory according to Eq.(21), and that wave packet spreading arises  because of quantum diffusion. 
 \\
\\
\begin{figure}[htb]
\centerline{\includegraphics[width=13cm]{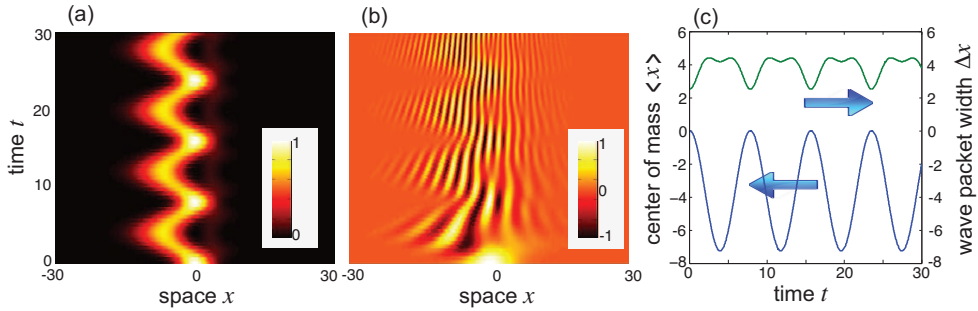}} \caption{Pseudo-periodic oscillations of a Gaussian wave packet for the generalized Wannier-Stark Hamiltonian (3) with $\epsilon=0$ and with a sinusoidal kinetic energy operator $T(\hat{p}_x)=\kappa \cos(d \hat{p}_x)$.  Evolution of (a) the probability density $|\psi(x,t)|^2$, and (b) of the real part of the wave function ${\rm Re}(\psi(x,t))$  on a pseudocolor map. (c) Evolution of the wave packet center of mass $\langle x(t) \rangle$ and width $\Delta x$. Parameter values are given in the text.}
\end{figure}
(ii){\it Second case: pseudo-Bloch oscillations.}  This case corresponds to the limit $\epsilon=0$ in Eq.(3) and displays a pseudo-periodic dynamics. In fact, at discretized times $t=t_m=mT_B$, integer multiplies than the BO period $T_B= 2 \pi /(Fd)$, from Eq.(20) one obtains
\begin{equation}
\mathcal{U}(x,y,t_m)= \exp(-iFxt_m) \delta(x-y)
\end{equation}
where we used the property $\sum_{n} \sigma_n \sigma_{n-l}^*= (1/F) \delta_{l,0}$.  Hence from Eqs.(14) and (24) on has
\begin{equation}
\psi(x,t_m)=\exp(-2 \pi i m x/d)  \psi(x,0).
\end{equation}
Equation (25) shows that the dynamics is pseudo-periodic, i.e. the probability distribution $|\psi(x,t)|^2$ undergoes a periodic dynamics with the BO period $T_B$, i.e. $|\psi(x,t_m)|^2=|\psi(x,0)|^2$, whereas the wave function $\psi(x,t)$ does not [this is due to the phase factor $\exp(-2 \pi i m x/d)$ in Eq.(25)]. An example of pseudo-periodic dynamics for an initial Gaussian wave packet is shown in Fig.2 for a sinusoidal shape of $T(q)$, i.e. $T(q)=\kappa \cos (qd)$. Parameter values used in the simulations are $\epsilon=0$, $F=0.2$, $\kappa=1$ and $d=4$, corresponding to a BO period $T_B= 2 \pi/(Fd) \simeq 7.85$. The initial condition is the Gaussian wave packet  $\psi(x,0) \propto \exp(-x^2/w^2)$ with $w=5$ as in Fig.1. Figure 2(a) and (b) depict on a pseudo color map the temporal evolution of the probability density $|\psi(x,t)|^2$ and of the real part ${\rm Re}(\psi(x,t))$ of the wave function, respectively. Note that, while the probability density undergoes a periodic oscillatory behavior analogous to the ordinary BOs in a crystal, the wave function does not; in particular the interference fringes visible in Fig.2(b) arise from the phase factor $\exp(-iFxt)$ appearing in Eq.(20).  Such a pseudo-periodic dynamical regime was previously introduced in Ref.\cite{r52} and referred to as {\it pseudo-Bloch oscillations}.\\
\\
\\
(iii) {\it Third-case: accelerated Bloch oscillations.} This is the most general case, corresponding to $T(q) \neq 0$ and $\epsilon \neq 0$. In this case the dynamical behavior is intermediate between the two previously considered regimes, i.e. one observes an oscillatory dynamics of the wave packet superimposed to the parabolic path as described by Eq.(21). Such a property can be proven by observing that, at times $t_m=mT_B$ ($m=0, \pm 1 , \pm 2,...$), the two propagators $\mathcal{U}(x,y,t)$ as given by Eq.(16) and Eq.(18) do coincide.  In fact, at times $t=t_m$ one has
\begin{equation}
G_l(t_m)=\sum_{n-\infty}^{\infty} \rho_n \rho^{*}_{n-l} = \Omega_l = \frac{1}{\epsilon^{2/3} F^{1/3}} \delta_{l,0}
\end{equation}
 where $\Omega_l$ are defined by Eq.(A.17) in Appendix A. Substitution of Eq.(26) into Eq.(16) yields
 \begin{equation}
 \mathcal{U}(x,y,t)=\sqrt{\frac{1}{4 \pi i \epsilon t_m}}  \exp \left( -iFxt_m -i \frac{\epsilon F^2 t_m^3}{3} \right)  \exp\left[ -\frac{\left( y-x- \epsilon F t_m^2 \right)^2}{4 i \epsilon t_m} \right] 
 \end{equation}
 which coincides with Eq.(18) taken at $t=t_m$. Indicating by $\hat{H}_1$ and $\hat{H}_2$ the two Hamiltonians defined by Eqs.(1) and (3), respectively, i.e. $\hat{H}_1$ is the limiting case of $\hat{H}_2$ when $T(q)=0$, the previous result can be formally written as \cite{note2}
 \begin{equation}
 \exp(-it_m \hat{H}_2)=\exp(-it_m \hat{H}_1)
 \end{equation}
 i.e. if the dynamics is mapped at discretized times $t_m$ integer multiplies than the BO period $T_B$ the two Hamiltonians (1) and (3) yields the same evolution. Therefore the oscillatory motion in each BO cycle is superimposed to the parabolic path (21). Such a dynamical behavior can be thus referred to as {\it accelerated Bloch oscillations}. At times different than $t_m=mT_B$, the following relation between the solutions $\psi_1(x,t)=\exp(-it \hat{H}_1) \psi(x,0)$ and $\psi_2(x,t)=\exp(-it \hat{H}_2) \psi(x,0)$ to the Schr\"{o}dinger equations with Hamiltonians $\hat{H}_1$ and $\hat{H}_2$ corresponding to the same initial condition $\psi(x,t)$ can be derived (see Appendix D)
 \begin{equation}
 \psi_2(x,t)= \sum_{l=-\infty}^{\infty} \Lambda_l(t) \psi_1(x-ld,t),
 \end{equation}
 where we have set
 \begin{equation}
 \Lambda_l(t)=F^{1/3} \epsilon^{2/3} G_l(t) \exp(-iFdlt)=F^{1/3} \epsilon^{2/3} \sum_{n=-\infty}^{\infty} \rho_n^* \rho_{n+l} \exp(iFdtn). \;\;\;\;
 \end{equation}
 For example, for a sinusoidal function $T(q)=\kappa \cos(qd)$ the explicit expressions of the functions $\Lambda_l(t)$, as obtained from Eqs.(30) and (C.5), read
 \begin{equation}
 \Lambda_l(t)= J_{-l} \left( \frac{\kappa}{Fd} \sqrt{2-2 \cos (Fdt) } \right) \left[ \frac{1-\exp(-iFdt)}{1-\exp(iFdt)}  \right]^{-l/2} \exp(-iFdlt). \;\;\;\;
 \end{equation}
 Equation (29) indicates that, at times $t \neq t_m$, the solution $\psi_2(x,t)$ is given by a suitable superposition (interference) of {\it shifted} replica of $\psi_1(x,t)$, weighted by the complex amplitudes $\Lambda_l(t)$. At $t=t_m$, $\Lambda_l(t_m)=\delta_{l,0}$ and thus $\psi_2(x,t_m)=\psi_1(x,t_m)$ as previously discussed.
\begin{figure}[htb]
\centerline{\includegraphics[width=11cm]{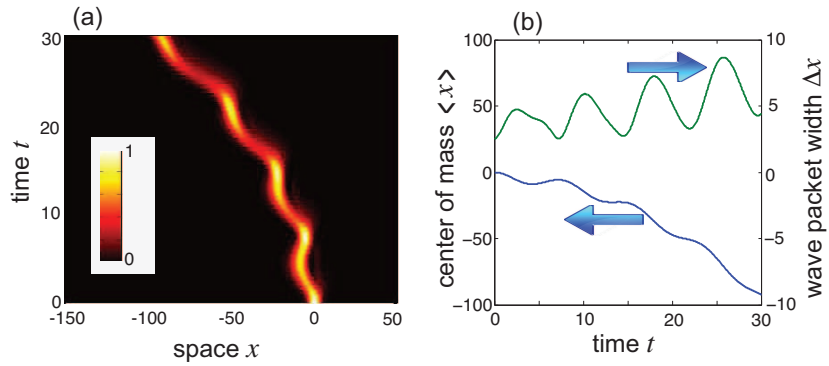}} \caption{Accelerated BOs of a Gaussian wave packet for the generalized Wannier-Stark Hamiltonian (3) with a sinusoidal kinetic energy operator $T(\hat{p}_x)=\kappa \cos(d \hat{p}_x)$. (a)  Evolution of the probability density $|\psi(x,t)|^2$ in a pseudo color map, and (b) of the wave packet center of mass and width. Parameter values are given in the text.}
\end{figure}
 
 An example of accelerated BOs is shown in Fig.3 for a sinusoidal function $T(q)=\kappa \cos(qd)$ and for parameter values $\epsilon=1/2$, $F=0.2$, $\kappa=1$ and $d=4$, i.e. for the same parameter values as in Fig.2 except for $\epsilon=1/2$.  The initial condition is the Gaussian wave packet  $\psi(x,0) \propto \exp(-x^2/w^2)$ with $w=5$ (i.e. as in Figs.1 and 2). The figure clearly shows that the wave packet undergoes and oscillatory motion with period $T_B$ over the averaged parabolic path of Fig.1. Note that, while in the pseudo BOs regime [case (ii), see Fig.2] wave packet spreading is suppressed, in the regime of accelerated BOs (Fig.3) the wave packet spreads on average following the same spreading law of the parabolic case (i) discussed above.

\section{Airy-Bloch oscillations}
As shown in Sec. 3.2, any normalizable wave packet  undergoes accelerated BOs and spreading when evolved by the generalized Wannier-Stark Hamiltonian (3) with a non-vanishing value of $\epsilon$. However, such a property can be violated by an initial wave packet that is not normalizable, i.e. such that $\int_{-\infty}^{\infty} dx \; |\psi(x,0)|^2= \infty$. An interesting case is the one corresponding to an initial condition $\psi(x,0)$ which is a generalized eigenfunction of the Stark Hamiltonian (1), i.e. an Airy wave packet. Let us assume, for example, the eigenfunction of (1) with energy $E=0$, which apart from a normalization constant is given by [see Eq.(10)]
\begin{equation}
\psi(x,0)={\rm Ai} \left( \left( \frac{F}{\epsilon} \right)^{1/3} x \right).
\end{equation}
For the free-particle Schr\"{o}dinger equation, i.e. for the Hamiltonian $\hat{H}=\epsilon \hat{p}_x^2$, Airy wave packets are shape preserving ones and undergo a self-accelerating motion, as originally shown by Berry and  Balazs in Ref.\cite{berry} and extended in several subsequent works (see, for example, \cite{AP3} and references therein). For the Stark Hamiltonian (1), they do not accelerate, i.e. they are at rest, because of the additional force $F$: indeed they are eigenstates of the Hamiltonian (1). Interestingly, we show now that for the generalized Wannier-Stark Hamiltonian (3) the Airy wave packet (32) evolves undergoing a {\it periodic} breathing dynamics with the BO period $T_B= 2 \pi/(Fd)$. Such a periodic and acceleration-free  breathing dynamics is referred to as {\it Airy-Bloch oscillations}. In fact, the evolution of the wave packet (32) under the generalized Wannier-Stark Hamiltonian (3) can be readily calculated using Eq.(29) and taking into account that $\psi(x,0)$ is an eigenfunction of (1) with zero energy. One obtains
\begin{equation}
\psi(x,t)=\sum_{l=-\infty}^{\infty} \Lambda_l(t) {\rm Ai} \left( \left( \frac{F}{\epsilon} \right)^{1/3} (x-ld) \right).
\end{equation}
Since $\Lambda_{l}(t_m)=\delta_{l,0}$ for $t_m= m T_B$, one has $\psi(x,t_m)=\psi(x,0)$, i.e. a strict periodic dynamics is obtained. Hence for the initial Airy distribution (32) of the wave function the acceleration motion is suppressed, and a periodic breathing dynamics is established.
An example of Airy-Bloch oscillations is shown in Fig.4(a). The periodic dynamics shown in the figure resembles a {\it quantum carpet} \cite{carpet} in Hamiltonians with absolutely continuous spectrum, such as quantum carpets arising from the Talbot effect for the free-particle Hamiltonian $\hat{H}=\epsilon  \hat{p}_x^2$ \cite{carpet,robi}.  
\begin{figure}[htb]
\centerline{\includegraphics[width=14cm]{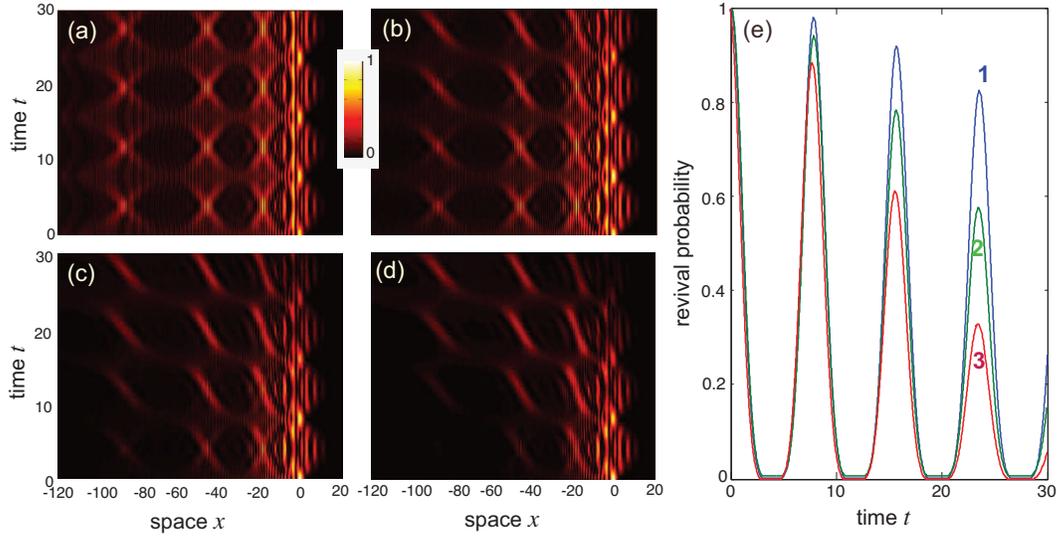}} \caption{Airy-Bloch oscillations. (a) Quantum carpet, showing the evolution of the probability density $|\psi(x,t)|^2$ in a pseudo color map for the initial condition (32) (ideal Airy wave packet). Parameter values are $\epsilon=1/2$, $F=0.2$, $T(q)=\kappa \cos(qd)$ with $d=4$ and $\kappa=1$. (b-c): same as (a), but for the initial wave packet distribution (34) and for increasing values of $a$: $a=300$ in (b), $a=100$ in (c) and $a=50$ in (d).
(e)  Evolution of the revival probability [Eq.(35)]. Curves 1,2 and 3 refer to the simulations in panels (b), (c) and (d), respectively.}
\end{figure}

In practice, the Airy distribution is an idealized one and corresponds to a delocalized (non-normalizable) state of the particle. However, it can be approximated by normalizable distributions obtained by enveloping the Airy function with a sufficiently-decaying function at $x \rightarrow -\infty$. For example, an initial normalizable wave packet distribution that approximates Eq.(32) is given by
\begin{equation}
\psi(x,0)=\mathcal{N} \exp(x/a) {\rm Ai} \left( \left( \frac{F}{\epsilon} \right)^{1/3} x \right),
\end{equation}
where $a>0$ and $\mathcal{N}$ is a normalization constant. Note that the function defined by Eq.(34) goes to zero at both $ x \rightarrow \pm \infty$ and is normalizable because the Airy function decays at $x \rightarrow \infty$ faster than exponential. Since the wave packet defined by Eq.(34) is normalizable, its center of mass undergoes an oscillatory and (on average) accelerated motion; however, for a sufficiently large values of $a$ the Airy-Bloch breathing dynamics of Fig.4(a) can be observed in the earlier BO cycles, as shown in Fig.4(b-d). After a few BO cycles, the periodicity is lost according to the analysis of Sec.3.2. This can be seen by computing the temporal evolution of the revival probability into the original state, defined as
\begin{equation}
P_{rev}(t)= \left|  \int_{-\infty}^{\infty}dx \; \psi^*(x,0) \psi(x,t) \right|^2.
\end{equation}
The behavior of $P_{rev}(t)$ is shown in Fg.4(e). Note that, as $a$ decreases, periodicity of the dynamics is rapidly lost.

\section{Conclusion}
In this work the phenomena of accelerated and Airy-Bloch oscillations have been predicted, which provide significant extensions of the famous Bloch oscillations 
originally predicted for electrons in a crystal under a uniform electric field. We introduced an exactly-solvable generalized Wannier-Stark Hamiltonian and showed rather generally that a dynamical regime intermediate between a pure oscillatory and an uniformly accelerated motion arises (accelerated Bloch oscillations). As a special case, for wave packets with an Airy shape  acceleration can be suppressed and a pure oscillatory (breathing) dynamics, leading to quantum carpets, is predicted  (Airy-Bloch oscillations). Owing to the possibility to emulate Schr\"{o}dinger equations with engineered  potentials and kinetic energy operators offered by optics \cite{uff1,uff2}, the predicted phenomena of accelerated and Airy-Bloch oscillations could be observed in an optical setting, as discussed in Ref.\cite{r52}.

%\begin{figure}[bt]
%\centerline{\psfig{file=ijmpbf1.eps,width=3.65in}}
%\vspace*{8pt}
%\caption{This is the caption for the figure. If the caption is 
%less than one line then it is centered. Long captions are 
%ustified to the full text width.}
%\end{figure}

\appendix{Generalized eigenfunctions of the Hamiltonian (3)}
In this Appendix we derive Eqs.(7) and (8), given in the text, for the improper eigenfunctions of the Hamiltonian (3). To this aim, let us first note that for $T(q)=0$ the eigenfunctions of $\hat{H}$ are simply given by shifted Airy functions (see e.g. \cite{r4}). For $T(q) \neq 0$, since $T(q)$ is periodic the action of the operator $T(\hat{p}_x)$ on a generic function $f(x)$ is to sum up shifted replica of the function, namely one has
\begin{equation}
T(\hat{p_x}) f(x)= \sum_n T_n \exp(nd \partial_x)f(x)=\sum_n T_n f(x+nd).
\end{equation}
 Hence we may look for a solution to the eigenvalue equation (6) of the form
 \begin{equation}
 \phi_E(x)= \sum_{n=-\infty}^{\infty} \rho_n {\rm Ai} \left( \alpha (x-nd-\beta) \right),
 \end{equation}
 where the coefficients $\rho_n$, $\alpha$ and $\beta$ in Eq.(A.2) are unknown parameters at this stage. Substitution of the Ansatz (A.2) into Eq.(6) and taking into account Eq.(A.1) and that $d^2 {\rm Ai} (x) /dx^2=x {\rm Ai}(x)$, one obtains
 \begin{eqnarray}
 (E-Fx) \sum_n \rho_n {\rm Ai} \left( \alpha (x-nd-\beta) \right)  =  -\epsilon \alpha^3 \sum_n (x-nd-\beta) \rho_n  {\rm Ai} \left( \alpha (x-nd-\beta) \right) \;\;\;\;\;\;\;\ \nonumber \\
  +  \sum_{l,n} T_l \rho_n {\rm Ai} \left( \alpha (x-nd+ld-\beta) \right). \;\;\;\;\;\;\;\;
 \end{eqnarray}
 To satisfy Eq.(A.3), the coefficients of the terms multiplying $x$ on the left and right hand sides of the equation must coincide. This yields the following expression for the 
 coefficient $\alpha$
 \begin{equation}
 \alpha= \left( \frac{F}{\epsilon} \right)^{1/3}
 \end{equation}
 and Eq.(A.3) simplifies as follows
 \begin{eqnarray}
 E\sum_n \rho_n {\rm Ai} \left( \alpha (x-nd-\beta) \right)  =  F \sum_n (nd+\beta)  \rho_n {\rm Ai} \left( \alpha (x-nd-\beta) \right) \nonumber \\
 + \sum_{n} \left( \sum_{l} T_l \rho_{n+l} \right)  {\rm Ai} \left( \alpha (x-nd-\beta) \right).
 \end{eqnarray}
 Equation (A.5) is satisfied provided that the following difference equation among the coefficients $\rho_n$ holds
 \begin{equation}
 (E -\beta F) \rho_n=Fdn \rho_n + \sum_l T_{l} \rho_{n+l}.
 \end{equation}
 To solve Eq.(A.6), let us introduce the  Fourier spectrum
 \begin{equation}
 S(q)=\sum_{-\infty}^{\infty} \rho_n \exp(-iqdn)
 \end{equation}
 which is a periodic function of $q$ with period $2 \pi/d$. The coefficients $\rho_n$ are determined from he spectrum $S(q)$ by the inverse relation
 \begin{equation}
 \rho_n= \frac{d}{2 \pi} \int_{-\pi/d}^{\pi/d} dq S(q) \exp(iqnd).
 \end{equation}
  From Eqs.(A.6) and (A.7) the following differential equation for the spectrum is obtained
  \begin{equation}
  i F \frac{dS}{dq}= \left[ E-\beta F -T(q) \right]S(q)
  \end{equation}
  which can be solved, yielding
  \begin{equation}
  S(q)=S(0) \exp \left\{ \frac{-i}{F}(E-\beta F)q+\frac{i}{F} \int_0^q d \xi T( \xi)  \right\}.
  \end{equation}
  The parameter $\beta$ is determined by imposing the periodicity of $S(q)$, i.e. that $S(q+2 \pi/d)=S(q)$. Since $\int_{0}^{2 \pi/d}dq T(q)=T_0=0$, the periodicity condition on $S(q)$ is ensured by assuming
  \begin{equation}
  \beta= E/F
  \end{equation}
 so that one has
 \begin{equation}
 S(q)= S(0) \exp \left\{\frac{i}{F} \int_0^q d \xi T( \xi)  \right\}.
 \end{equation}
 Substitution of Eq.(A.12) into Eq.(A.8) yields
 \begin{equation}
 \rho_n= \frac{d S(0)}{2 \pi} \int_{-\pi/d}^{\pi/d} dq \exp \left\{iqdn+ \frac{i}{F} \int_0^q d \xi T( \xi)  \right\}.
 \end{equation}
 Finally, the constant $S(0)$ is determined by imposing that the generalized eigenfunctions $\phi_E(x)$ satisfy the usual normalization condition (9) given in the text. To this aim, let us explicitly calculate the scalar product $\langle \phi_{E'}(x) | \phi_E(x) \rangle$ using Eq.(A.2). One obtains
 \begin{equation}
\langle \phi_{E'}(x) | \phi_E(x) \rangle= \sum_{n,l} \rho_n^* \rho_l \langle {\rm Ai}(\alpha(x-nd-E'/F) | {\rm Ai} (x-ld-E/F) \rangle. \;\;\;\;
 \end{equation}
  Taking into account that
  \begin{equation}
  \int_{-\infty}^{\infty} {\rm } dx {\rm Ai} [\alpha(x-a)] {\rm Ai} [\alpha(x-b)]= \frac{1}{\alpha^2} \delta(a-b)
  \end{equation}
 one has
 \begin{eqnarray}
\langle \phi_{E'}(x) | \phi_E(x) \rangle= \frac{F}{\alpha^2}\sum_{n,l} \rho_n^* \rho_l \delta (E'-E+Fdn-Fdl) =  \;\;\;\; \\
= \frac{F}{\alpha^2} \sum_{ l} \Omega_l \delta (E'-E-Fdl). \;\;\; \nonumber
 \end{eqnarray}
 where we have set
 \begin{equation}
\Omega_{l} \equiv  \sum_n \rho_n^* \rho_{n+l} 
 \end{equation}
 Using Eq.(A.7) it can be readily shown that
 \begin{equation}
 \Omega_{l}= \frac{d}{2 \pi} \int_{-\pi/d}^{\pi /d} dq |S(q)|^2 \exp(iql).
 \end{equation}
 Since $|S(q)|^2=|S(0)|^2$ [see Eq.(A.12)], one has $\Omega_l=|S(0)|^2 \delta_{l,0}$, and thus
 \begin{equation}
 \langle \phi_{E'}(x) | \phi_E(x) \rangle= \frac{|S(0)|^2 F}{\alpha^2} \delta(E'-E).
 \end{equation}
 Hence the normalization (9) is obtained by assuming
 \begin{equation}
 S(0)=\frac{\alpha}{ \sqrt{F}}= \frac{1}{\epsilon^{1/3} F^{1/6}}.
 \end{equation}
 Substitution of Eq.(A.20) into Eq.(A.13) yields the expression (8) given in the text for the coefficients $\rho_n$.
 
\appendix{Propagator of the Hamiltonian (3)}
In this Appendix we derive the analytical form (16) given in the text for the kernel $\mathcal{U}(x,y,t)$ of the propagator $\hat{U}(t)$ of the Hamiltonian (3). Substitution of Eq.(7) into Eq.(15) yields
{\small
\begin{equation}
\mathcal{U}(x,y,t)=\sum_{n,l} \rho_l^* \rho_n \int_{-\infty}^{\infty} dE  {\rm Ai} \left(\alpha (y-ld-E/F) \right) {\rm Ai} \left( \alpha (x-nd-E/F) \right) \exp(-i Et) \;\;\;\;\;\;
\end{equation}}
\noindent with $\alpha=(F/ \epsilon)^{1/3}$.  After a change of the integration variable and summation indices on the right hand side of Eq.(B.1) one obtains
\begin{equation}
\mathcal{U}(x,y,t)=\frac{F}{\alpha} \exp(-iFtx) \sum_l \left( \sum_n \rho_n \rho_{n-l}^* \exp(iFtdn) \right) \Phi(y-x+dl,t) \;\;\;\;
\end{equation}
where we have set
\begin{equation}
\Phi(x,t) \equiv \int_{-\infty}^{\infty} d \xi {\rm Ai} (\xi) {\rm Ai}(\xi+ \alpha x) \exp(iFt \xi / \alpha).
\end{equation}
The integral of the product of Airy functions on the right hand side of Eq.(B.3) with $\alpha$ given by Eq.(A.4) can be calculated in a close form and reads
\begin{equation}
\Phi(x,t)=\sqrt{\frac{1}{4 \pi i \epsilon t \alpha^2}} \exp\left[ -\frac{\left( x- \epsilon F t^2 \right)^2}{4 i \epsilon t} - \frac{i \epsilon F^2 t^3}{3}\right]
\end{equation}
After setting
\begin{equation}
G_l(t) \equiv \sum_n \rho_n \rho_{n-l}^* \exp(iFtdn) 
\end{equation}
one then obtains
{\small
\begin{equation}
\mathcal{U}(x,y,t)=\frac{F}{\alpha^2} \sqrt{\frac{1}{4 \pi i \epsilon t}}  \exp \left( -iFxt -i \frac{\epsilon F^2 t^3}{3} \right) \sum_{l= -\infty}^{\infty}  G_l(t) \exp\left[ -\frac{\left( y-x+dl- \epsilon F t^2 \right)^2}{4 i \epsilon t} \right] \;\;
\end{equation}}
\noindent 
From Eqs.(A.4) and (B.6) then follows Eq.(16) given in the text. 
 
 \appendix{Propagator for a sinusoidal lattice band}
Let us consider the case of a sinusoidal shape for the function $T(q)$, namely
\begin{equation}
T(q)= \kappa \cos (qd)
\end{equation}
 which corresponds in ordinary theory of electronic states in crystals to the well-known tight-binding lattice model in the nearest-neighbor approximation. In this case from Eq.(8) one obtains
 \begin{equation}
 \rho_n=\frac{1}{\epsilon^{1/3} F^{1/6}} J_n \left( -\frac{\kappa}{Fd} \right).
 \end{equation}
 In deriving Eq.(C.2) the following integral representation of Bessel $J_n$ function has been used
 \begin{equation}
 J_n(y)=\frac{1}{2 \pi} \int_{-\pi}^{\pi} dx \exp (inx-iy \sin x).
 \end{equation}
 The functions $G_l(y)$, given by Eq.(17), take the form
 \begin{equation}
G_l(t) = \frac{1}{\epsilon^{2/3} F^{1/3}}  \sum_{n=-\infty}^{\infty} J_n \left( -\frac{\kappa}{Fd} \right) J_{n-l} \left( -\frac{\kappa}{Fd} \right)  \exp(iFtdn).
\end{equation}
 The sum on the right hand side of Eq.(C.4) can be calculated using the Graf's addition formula for Bessel functions ( cf. Sec.11.3 (1) of Ref.\cite{rapp}), yielding
 \begin{equation}
 G_l(t)=\frac{1}{\epsilon^{2/3} F^{1/3}} J_{-l} \left( \frac{\kappa}{Fd} \sqrt{2-2 \cos (Fdt) } \right) \left[ \frac{1-\exp(-iFdt)}{1-\exp(iFdt)}  \right]^{-l/2}.
 \end{equation}
 Finally, the propagator (16) takes the analytical form
 {\small
\begin{eqnarray}
\mathcal{U}(x,y,t) =\sqrt{\frac{1}{4 \pi i \epsilon t}}  \exp \left( -iFxt -i \frac{\epsilon F^2 t^3}{3} \right) \sum_{l= -\infty}^{\infty}  J_{l} \left( \frac{\kappa}{Fd} \sqrt{2-2 \cos (Fdt) } \right) \times \nonumber \\
\times  \left[ \frac{1-\exp(-iFdt)}{1-\exp(iFdt)}  \right]^{l/2}
\exp\left[ -\frac{\left( y-x-dl- \epsilon F t^2 \right)^2}{4 i \epsilon t} \right] \;\;
\end{eqnarray}}
 \section{Proof of Eq.(29)}
 Let us indicate by 
 \begin{equation}
 \psi_1(x,t)= \exp(-it \hat{H}_1) \psi(x,0) \;\; , \;\;\;  \psi_2(x,t)= \exp(-it \hat{H}_2) \psi(x,0) 
 \end{equation}
 the solutions to the Schr\"{o}dinger equation $ i \partial_t \psi=\hat{H} \psi$ with Hamiltonians $\hat{H}_1=\epsilon \hat{p}_x^2+Fx$ and $\hat{H}_2=\hat{H}_1+T( \hat{p}_x)$, respectively, corresponding to the same initial condition $\psi(x,0)$. One has
 \begin{equation}
 \psi_1(x,t)=\int_{-\infty}^{\infty} dy\;  \mathcal{U}_1(x,y,t) \psi(y,0) 
 \end{equation}
 \begin{equation}
   \psi_2(x,t)=\int_{-\infty}^{\infty} dy\;  \mathcal{U}_2(x,y,t) \psi(y,0)
 \end{equation}
 where the propagators $\mathcal{U}_1(x,y,t)$ and $\mathcal{U}_2(x,y,t)$ are given by Eqs.(18) and (16), respectively. From a comparison of Eqs.(16) and (18) one can readily write
 \begin{equation}
 \mathcal{U}_2(x,y,t)= \sum_{l=-\infty}^{\infty} \Lambda_l(t) \mathcal{U}_1(x-ld,y,t)
 \end{equation}
 where the functions $\Lambda_l(t)$ are defined by Eq.(30) given in the text. Substitution of Eq.(D.4) into Eq.(D.3) finally yields Eq.(29) given in the text.


\begin{thebibliography}{0}
\bibitem{note1}
 Such a result is exact and readily follows from the Eherenfest theorem.
\bibitem{note2}
 Such an equivalence could be also proven by operatorial methods using the Zassenhau and Baker-Campbell-Hausdorff formulas.
\end{thebibliography}
\end{document}